\documentclass[aps,prb,amsmath,preprint,showpacs,superscriptaddress]{revtex4}
\usepackage{graphicx}

\begin{document}

\title{Optical probing of spin fluctuations of a single magnetic atom}

\author{L. Besombes}
\email{lucien.besombes@grenoble.cnrs.fr}
\author{Y. Leger}
\author{J. Bernos}
\author{H. Boukari}
\author{H. Mariette}
\author{J.P. Poizat}
\affiliation{CEA-CNRS group "Nanophysique et Semiconducteurs",
Institut N\'eel, CNRS \& Universit\'e Joseph Fourier, 25 avenue des
Martyrs, 38042 Grenoble, France}
\author{J. Fern\'andez-Rossier}
\affiliation{Departamento de F\'{\i}sica Aplicada, Universidad de
Alicante, San Vicente del Raspeig, 03690 Spain}
\author{R. Aguado}
\affiliation{Instituto de Ciencia de Materiales de Madrid, CSIC,
Madrid, Spain}


\begin{abstract}

We analyzed the photoluminescence intermittency generated by a
single paramagnetic spin localized in an individual semiconductor
quantum dot. The statistics of the photons emitted by the quantum
dot reflect the quantum fluctuations of the localized spin
interacting with the injected carriers. Photon correlation
measurements which are reported here reveal unique signatures of
these fluctuations. A phenomenological model is proposed to
quantitatively describe these observations, allowing a measurement
of the spin dynamics of an individual magnetic atom at zero magnetic
field. These results demonstrate the existence of an efficient spin
relaxation channel arising from a spin-exchange with individual
carriers surrounding the quantum dot. A theoretical description of a
spin-flip mechanism involving spin exchange with surrounding
carriers gives relaxation times in good agreement with the measured
dynamics.

\end{abstract}


\maketitle

\section{Introduction}
The decrease of the structure size in semiconductor electronic
devices and magnetic information storage devices has dramatically
reduced the number of atoms necessary to process and store bit of
information. Information storage on a single magnetic atom would be
an ultimate limit. The performance of such memory elements will be
governed by the quantum fluctuations of the localized spins
\cite{xiao2004}. Diluted magnetic semiconductors (DMS) systems
combining high quality semiconductor structures and the magnetic
properties of Mn impurities are good candidates for these ultimate
spintronic devices \cite{Fernandez2007}. It has been shown that in a
DMS with low magnetic atom concentration, the spin dynamics under
magnetic field is ultimately controlled by the spin-lattice coupling
\cite{lambe60,kneip06}. An extrapolation of the spin dynamics
measurements in bulk DMS suggests that a long spin relaxation time
in the millisecond range could be expected for an isolated Mn spin
\cite{dietl95}. However, despite the recent development of different
successful techniques to address a single spin
\cite{heinrich2004,Kitchen2006}, such dynamics has never been
directly observed.

Growth and optical addressing of DMS quantum dots (QDs) containing a
single magnetic atom were achieved recently
\cite{Besombes04,Maingault2006}. When a Mn dopant atom is included
in a II-VI QD, the spin of the optically created electron-hole pair
interacts with the five {\it d} electrons of the Mn (total spin
M=5/2). This leads to a splitting of the once simple
photoluminescence (PL) spectrum of an individual QD into six (2M+1)
components. This splitting results from the spin structure of the
confined holes which are quantized along the QDs' growth axis with
their spin component taking only the values J$_z$=$\pm$3/2. The
hole-Mn exchange interaction reduces to an Ising term J$_z$.M$_z$
and shifts the emission energy of the QD, depending on the relative
projection of the Mn and hole spins. As the spin state of the Mn
atom fluctuates during the optical measurements, the six lines are
observed simultaneously in time average PL spectrum. The intensities
of the lines reflect the probability for the Mn to be in one of its
six spin components and the emitted photon is a probe of the spin
state of the Mn when the exciton recombines.

\begin{figure}
[bt]
\includegraphics[width=2.8in]{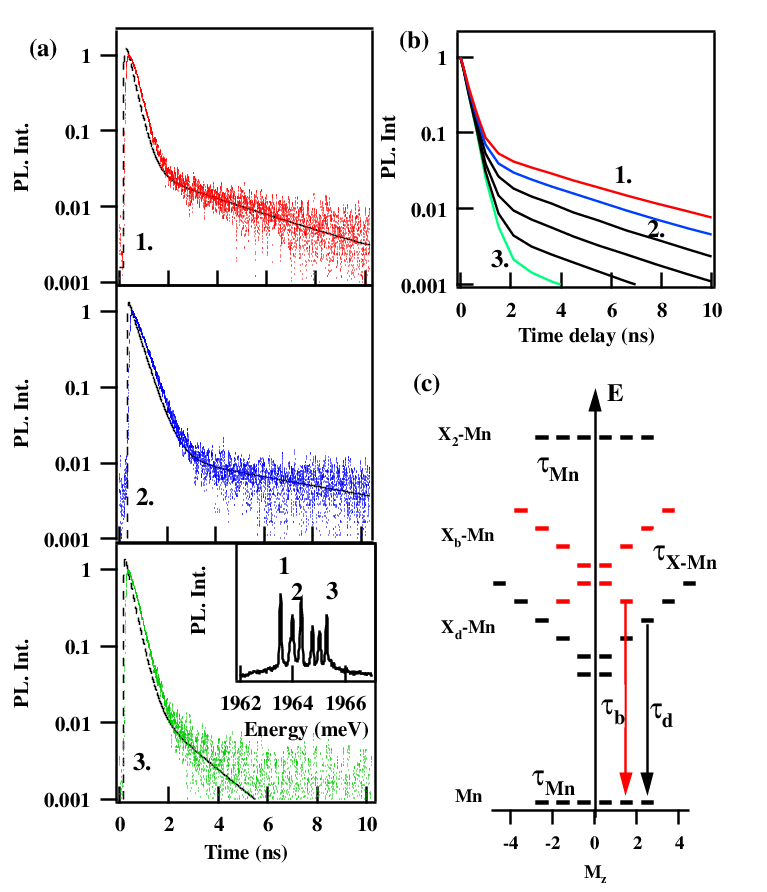}
\caption{ (Color online) (a) Experimental time resolved PL spectra
recorded on three different energy lines (labeled 1, 2 and 3) of an
X-Mn complex. The inset shows the corresponding time integrated PL
spectrum. (b) PL decay time calculated using the parameters T=5K,
$\tau_{b}$=280ps, $\tau_{d}$=8ns, $\tau_{X-Mn}$=25ns. (c) Energy
levels involved in the rate equation model described in the text
displayed as a function of their total angular momentum S$_z$.}
\label{fig1}
\end{figure}

In this article, we show how we can use the statistics of the
photons emitted by a single Mn doped QD to probe the spin dynamics
of a magnetic atom interacting with both confined and free carriers.
We performed correlations of the PL intensity emitted by individual
lines of an isolated Mn-doped QD. In these start-stop experiments
\cite{Couteau2004}, the detection of the first photon indicates by
its energy and polarisation that the Mn spin has a given
orientation. The detection probability of a second photon with the
same energy and polarisation is then proportional to the probability
of conserving this spin state. The time evolution of this intensity
correlation signal is a probe of the spin dynamics of the Mn atom.
The auto-correlation signal displays a bunching effect revealing a
PL intermittency. This intermittency results from fluctuations of
the Mn spin. Correlation of the intensity emitted by opposite spin
states of the exciton-Mn complex (namely,cross-correlation) presents
an antibunching at short delays. The characteristic time of this
antibunching corresponds to the spin transfer time between the two
degenerated spin states. A thermalization on the exciton-Mn complex
is directly evidence by the energy and temperature dependences of
the correlation curves. The measured single Mn spin relaxation times
are in the range of 20 ns and are strongly influenced by the
injection of carriers in the vicinity of the QDs. Mn spin-flips
induced by the injection of carriers in and around the QD are
theoretically described. These scattering processes give relaxation
times in good agreement with the measured dynamics.

\section{Samples and experiments}

Single Mn atoms are introduced in CdTe/ZnTe QDs during their growth
by molecular beam epitaxy adjusting the density of Mn atoms to be
roughly equal to the density of QDs \cite{Maingault2006}. The
statistics of the photons emitted by individual Mn-doped QDs was
analyzed using a combination of a low-temperature (5K) scanning
optical microscope and a Hanbury Brown and Twiss (HBT) setup for
photon-correlation measurements \cite{Couteau2004}. High refractive
index hemispherical solid immersion lens were used to enhance the
collection of the single dot emission. The PL was quasi-resonantly
excited with a tunable CW dye laser or by picosecond pulses from a
doubled optical parametric oscillator and collected through aluminum
shadow masks. The circularly polarized collected light was
spectrally dispersed by a 1 m monochromator before being detected in
the HBT setup or by a single fast avalanche photodiode (time
resolution 40 ps) for time resolved measurements. The time
resolution of the HBT setup was about 500 ps. Under our experimental
conditions with counts rates of a few kHz the measured photon pair
distribution yields after normalization the intensity
autocorrelation function g$^{(2)}$($\tau$).

\section{Exciton-Mn spin flips.}

\begin{figure}[bt]
\includegraphics[width=4in]{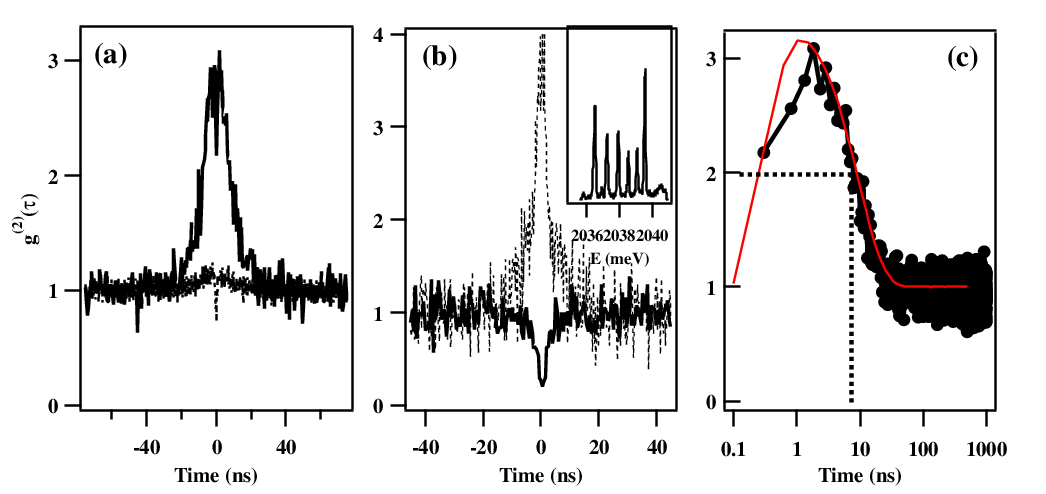}
\caption{(a) Auto-correlation function of the intensity collected in
$\sigma+$ polarisation on the low energy line of the X-Mn complex
(solid line) and on the overall PL spectra (dotted line). (b)
Circularly polarized cross-correlation function (solid line) and
auto-correlation (dotted line) on the same line as in (a) but for a
larger excitation intensity. The inset shows the PL spectrum of the
corresponding QD. (c) Experimental auto-correlation function and
theoretical function calculated with the rate equation model
described in the text with parameters T=5K, $\tau_{b}$=280ps,
$\tau_{d}$=8ns, $\tau_{X-Mn}$=25ns, $\tau_{Mn}$=50ns and
g=0.05x10$^{-3}$ps$^{-1}$.} \label{fig2}
\end{figure}

A signature of the Mn spin dynamics can be observed first in the PL
decay of the X-Mn complex. Fig.1 presents the PL decay of three
different transitions of the X-Mn complex. These transitions present
a biexponential decay. The fast component corresponds to the
radiative lifetime of the exciton, as already measured in non
magnetic QDs. The long component is associated with the existence of
the dark excitons \cite{smith05}. Direct recombination of the dark
exciton can be observed in some Mn-doped QDs because of a slight
admixture of the bright states with the dark ones induced by a
valence band mixing. However, the dark excitons mainly contribute to
the signal by undergoing a spin flip to become bright excitons which
decays radiatively. The PL decay is then determined by both
radiative decay and excitons spin-flips. The exciton decay, and
particularly the amplitude of the slow component, depends strongly
on the energy level observed. For the high energy lines, the slow
component makes a negligible contribution to the time integrated
signal. Conversely, for the low energy lines, the secondary
component makes a significant contribution while the primary
lifetime remains constant. In this regime, the secondary lifetime
can be associated either with the dark exciton lifetime or the
exciton spin flip time.

To extract these two parameters from the PL decay curves, we compare
the experimental data with a rate equation model describing the time
evolution of the population of the 24 X-Mn spin levels (Fig.1(c))
after the injection of a single exciton \cite{govorov2005}.
Different spin-flips times are expected depending on wether the
transitions occur with or without conservation of the energy or of
the total spin. However, we consider in first approximation that the
spin-flips among the X-Mn states can be described by a single
characteristic time $\tau_{X-Mn}$. We consider that at finite
temperature, the intraband relaxation rates
$\Gamma_{\gamma\rightarrow\gamma'}$ between the different spin
states of the exciton-Mn complex depend on the energy separation
$E_{\gamma\gamma'}=E_{\gamma'}-E_{\gamma}$. Here we use
$\Gamma_{\gamma\rightarrow\gamma'}$=1/$\tau_{X-Mn}$ if
$E_{\gamma\gamma'}<0$ and
$\Gamma_{\gamma\rightarrow\gamma'}$=1/$\tau_{X-Mn}e^{-E_{\gamma\gamma'}/k_BT}$
if $E_{\gamma\gamma'}>0$ \cite{govorov2005}. This describes a
partial thermalization among the 24 levels of the X-Mn system during
the lifetime of the exciton (bright or dark). In this model, we also
neglect the influence of the valence band mixing on the oscillator
strength \cite{Leger2007} and consider that all the excitonic bright
(dark) states have the same lifetime $\tau_b$ ($\tau_d$). Because of
an efficient hole spin-flip during the phonon assisted relaxation of
the unpolarized injected carriers, we consider that the excitons
with spins $\pm1$ and $\pm2$ are excited with the same probability
at $t=0$. Only optical transitions conserving the Mn spin are
considered. The calculation shows that changing $\tau_d$ mainly
influences the characteristic time of the long decay component
whereas changing $\tau_{X-Mn}$ affects also the amplitude of this
component. The PL decay curves can be reproduced well by this rate
equation model using $\tau_b=280ps$, $\tau_d=8ns$ and
$\tau_{X-Mn}=25ns$ (Fig. 1(b)). The value of $\tau_d$ controls the
decay time of the long component whereas $\tau_{X-Mn}$, larger than
$\tau_d$, reproduces the emission energy dependence of the amplitude
of the long component very well.

Within the X-Mn complex, the relaxation time between different spin
states $\tau_{X-Mn}$ can be affected by the spin orbit,
carrier-phonon and exchange interactions \cite{govorov2005} that
affect the exciton. In QDs, the electron spin relaxation is longer
than the radiative lifetime and is ultimately controlled by random
fluctuations of the nuclear spins \cite{feng2007}. The hole spin
relaxation is mainly controlled by the interaction with phonons
\cite{woods2004} and can be faster explaining the observed partial
thermalization of the X-Mn complex during the lifetime of the ground
state exciton.

\section{Fluctuations of an isolated Mn spin.}

To directly observe the time fluctuations of the Mn spin interacting
with the injected carriers, we analyzed the statistics of the
photons emitted by a Mn-doped QD. This statistics can be deduced
from an intensity correlation function of the QD emission. Fig. 2(a)
shows the intensity correlation function g$^{(2)}$($\tau$) of the
circularly polarized ($\sigma+$) low energy line of a Mn-doped QD
compared with the correlation function obtained for the overall PL
of the QD. The auto-correlation function obtained for all the
photons emitted by the QD is characteristic of a single photon
emitter with a dip at short delay. The width of this antibunching
signal is given by the lifetime of the emitter and the generation
rate of excitons and its depth is limited by the time resolution of
the HBT setup. A similar experiment performed on one of the single
line of the X-Mn complex still presents a reduced coincidence rate
near $\tau$=0, but it is mainly characterized by a large bunching
signal with a half width at half maximum (HWHM) of about 10ns. This
bunching reflects an intermittency in the QD emission. This
intermittency likely comes from the fluctuations of the Mn spin
orientation.

To confirm this result, cross-correlation measurements were
performed between different spin states of the X-Mn complex.
Cross-correlation of the $\sigma+$ and $\sigma-$ photon emitted by
the low energy line (fig.2(b)) shows an antibunching with
g$^{(2)}$(0)=0.2 and a HWHM of about 5ns. These two different
behaviors, namely the bunching of the auto-correlation signal and
the antibunching of the cross-correlation signal, demonstrate
unambiguously that the statistic of the QD emission is completely
governed by the Mn spin fluctuations. Whereas the auto-correlation
probes the time dependence of the probability for the spin of the Mn
to be conserved (M$_z$ =+5/2 at $\tau$=0 in Fig.2(a)), the
cross-correlation presented in Fig.2(b) is a probe of the spin
transfer between +5/2 and -5/2.

\begin{figure}[bt]
\includegraphics[width=2.8in]{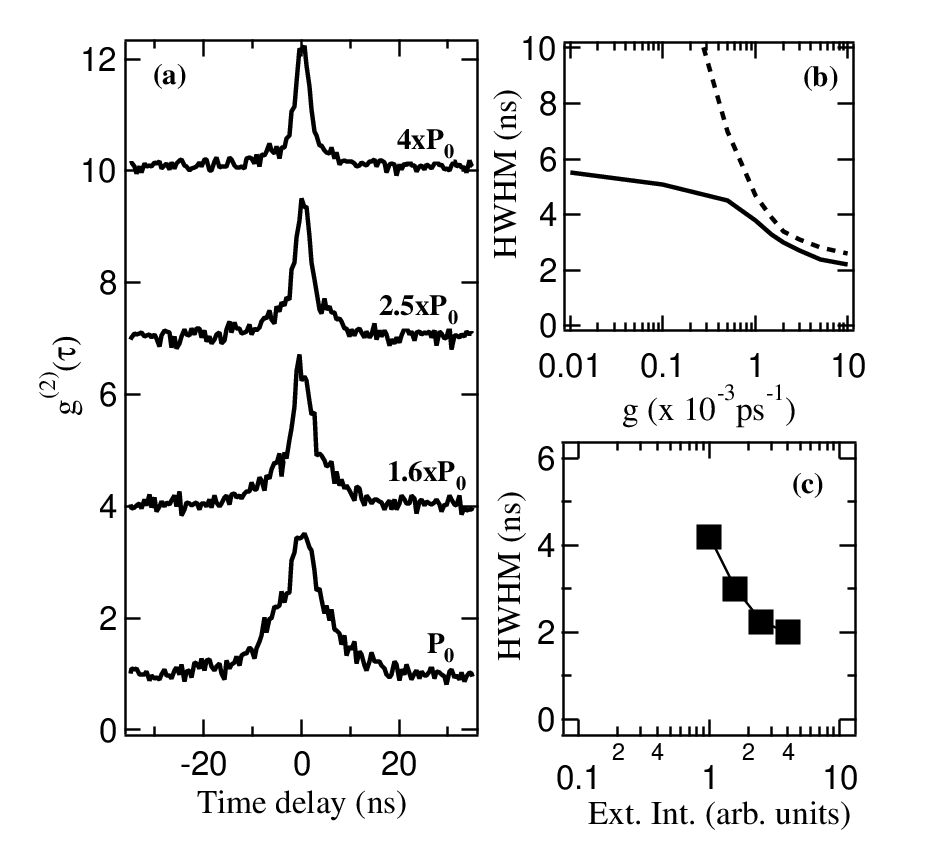}
\caption{ (Color online) (a) Power dependence of the autocorrelation
function of the hight energy line of an X-Mn complex.  (b)
Calculated power dependence of the HWHM of the autocorrelation
function of the hight energy line. The parameters used in the model
are: $\tau_{b}$=280ps, $\tau_{d}$=8ns, $\tau_{X-Mn}$=20ns, and
$\tau_{Mn}$=40ns (plain line) or  $\tau_{Mn}$=4$\mu$s (dotted line).
(c) Experimental HWHM of the autocorrelation signals presented in
(a).} \label{fig3}
\end{figure}

As observed in the time resolved PL measurements, fluctuations of
the Mn spin occur during the lifetime of an exciton. However, they
can also take place when the QD is empty. As the spin relaxation
rate of the Mn is expected to be influenced by the presence of
carriers in the QD, we have to consider two relaxation times,
$\tau_{Mn}$ for an empty dot and $\tau_{X-Mn}$ for a dot occupied by
an exciton. Their relative contributions to the observed effective
relaxation time will depend on the generation rate of excitons. The
rate equation model described previously can be extended to extract
an order of magnitude of the parameters $\tau_{Mn}$ and
$\tau_{X-Mn}$ from the correlation experiments. Six biexciton states
are added to the 24 $(X-Mn)$+6 $(Mn)$ level system (see Fig.1(c)). A
continuous generation rate $g$ is considered to populate the exciton
and biexciton states. The initial state of the system is fixed on
one of the six Mn ground states and one monitors the time evolution
of the population of the corresponding bright X-Mn state. When
normalized to one at long time, this time evolution accounts for the
correlation function of the transition associated with the
considered X-Mn level.

The time evolution of the correlation function calculated using this
model are presented in Fig.2(c) and compared with the experimental
data. At low generation rate, when the average time between two
injected excitons is longer than any spin relaxation rate,
$\tau_{Mn}$ and $\tau_{X-Mn}$ have distinguishable effects on the
calculated correlation curves. In average, the relaxation of the Mn
alone (controlled by $\tau_{Mn}$) is only perturbed by the injection
of the exciton used to probe the Mn spin projection. During the
lifetime of this exciton, the system relaxes with the relaxation
rate $\tau_{X-Mn}$. This produces a relaxation of the Mn spin
proportional to the ratio of $\tau_{X-Mn}$ and the exciton lifetime.
A reduction of $\tau_{X-Mn}$ then reduces the amplitude of the
bunching curve expected for a Mn alone (because of the six available
spin states, the maximum amplitude of the bunching should be six)
without significantly changing its width controlled by $\tau_{Mn}$.
With the generation rate used in the measurements of Fig.2(a) (a
generation rate of about g=0.05x10$^{-3}$ps$^{-1}$ can be deduced
from the ratio of the exciton and biexciton amplitude),
$\tau_{X-Mn}$ mainly affects the height of the bunching signal
whereas $\tau_{Mn}$ preferentially controls its width. Then, with
given values of $g$, $\tau_b$ and $\tau_d$, it is possible to
extract a parameter pair ($\tau_{Mn}$,$\tau_{X-Mn}$) that reproduces
the bunching and anti-bunching curves. The bright and dark excitons
lifetimes were estimated from the PL decay curves and the exciton
generation rate can be estimated from the relative amplitudes of the
exciton and biexciton emissions \cite{bes05}. For the data presented
in Fig.2(c), the best fit is obtained with $\tau_{X-Mn}=25ns$ and
$\tau_{Mn}=50ns$. The X-Mn relaxation time obtained is consistent
with the value deduced from the PL decay curves. The relaxation time
of the Mn alone (empty dot) appears to be 3 orders of magnitude
shorter than expected from the extrapolation of measurements in bulk
dilute CdMnTe under magnetic field \cite{dietl95}.

\section{Carriers induced Mn spin fluctuations}

The observed Mn spin dynamics is not simply an intrinsic property of
the localized Mn atom but depends on the optical excitation
conditions. The power dependence of the correlation signal of the
high energy transition of a X-Mn complex is presented in Fig.3(a).
Increasing the excitation power significantly reduces the width of
the correlation signal. This reduction has two origins: first, when
carriers are injected in the QD under quasi-resonant conditions
(excitation on an excited state of the QD), increasing the carrier
generation rate increases the probability of finding the QD occupied
by an exciton. The spin relaxation time being shorter for an
occupied dot than for an empty dot, the Mn spin fluctuates faster
and the width of the auto-correlation curve decreases. This effect
is illustrated by the power dependence of the HWHM of the calculated
and experimental correlation curves presented in Fig.3(b) and 3(c)
respectively. At high generation rate, the width of the correlation
signal is controlled by $\tau_{X-Mn}$ whereas at low excitation the
photon statistics is ultimately determine by the spin fluctuations
of the Mn alone. The width of the calculated correlation curves
saturates at low excitation. This maximum width is controlled by
$\tau_{Mn}$. In the experiments, this saturation is not observed due
to the limit in the accessible excitation power range.

\begin{figure}[bt]
\includegraphics[width=2.8in]{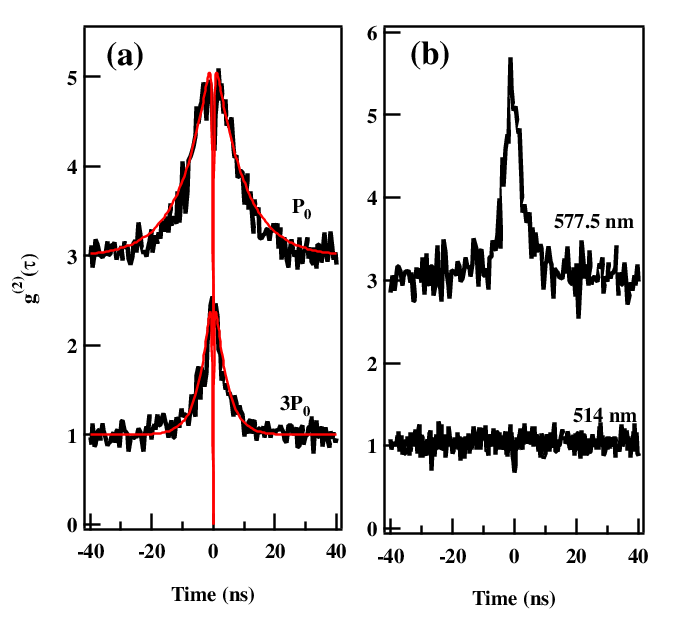}
\caption{ (color online)(a) Auto-correlation function on the low
energy line of an X-Mn complex in $\sigma+$ polarization for
excitation intensities P$_0$ and 3P$_0$. Theoretical curves are
presented in red. A reduction of $\tau_{X-Mn}$(=15ns) and
$\tau_{Mn}$(=20ns) has to be included to describe the high
excitation intensity autocorrelation curve.  (b) Auto-correlation
function on the low energy line of an X-Mn complex in $\sigma+$
polarization for two different excitation conditions: resonant on an
excited state (577.5nm) and non-resonant (514nm).} \label{fig4}
\end{figure}

However, this process is not sufficient to explain the observed
excitation power dependence of the correlation signal in all the
investigated QDs. For instance, to reproduce the power dependence
presented in Fig.4(a), one has also to reduce the spin relaxation
times $\tau_{Mn}$ and $\tau_{X-Mn}$ at high excitation intensity. In
Fig.4(a) the best fit at high excitation power is obtained with
$\tau_{Mn}=25$ and $\tau_{X-Mn}=15$ whereas at low excitation
$\tau_{Mn}=50$ and $\tau_{X-Mn}=25$. This reduction of the
relaxation time likely comes from the spin-spin coupling with
carriers injected in the surroundings of the QD thought the
background absorption observed in PLE spectra of this individual QD
\cite{vasanelli02}.

The influence of the free carriers on the spin relaxation rate is
shown by the correlation signals obtained on the same X-Mn
transition for two different excitation wavelengths (Fig.4(b)):
resonant on an excited state (577,5nm) and non-resonant in the ZnTe
barriers (514nm). These two signals are recorded with the same
photon count rate, suggesting a similar occupation rate of the QD.
The characteristic bunching signal observed under quasi-resonant
excitation completely disappears when the excitation energy is tuned
above the wetting layer absorption. As already observed in DMS
quantum wells, this behavior reflects the extreme sensitivity of the
localized Mn spin to the spin-spin coupling with the free carriers
or the carriers relaxing in the QD \cite{Tyazhlov97}.

\begin{figure}[bt]
\includegraphics[width=2.8in]{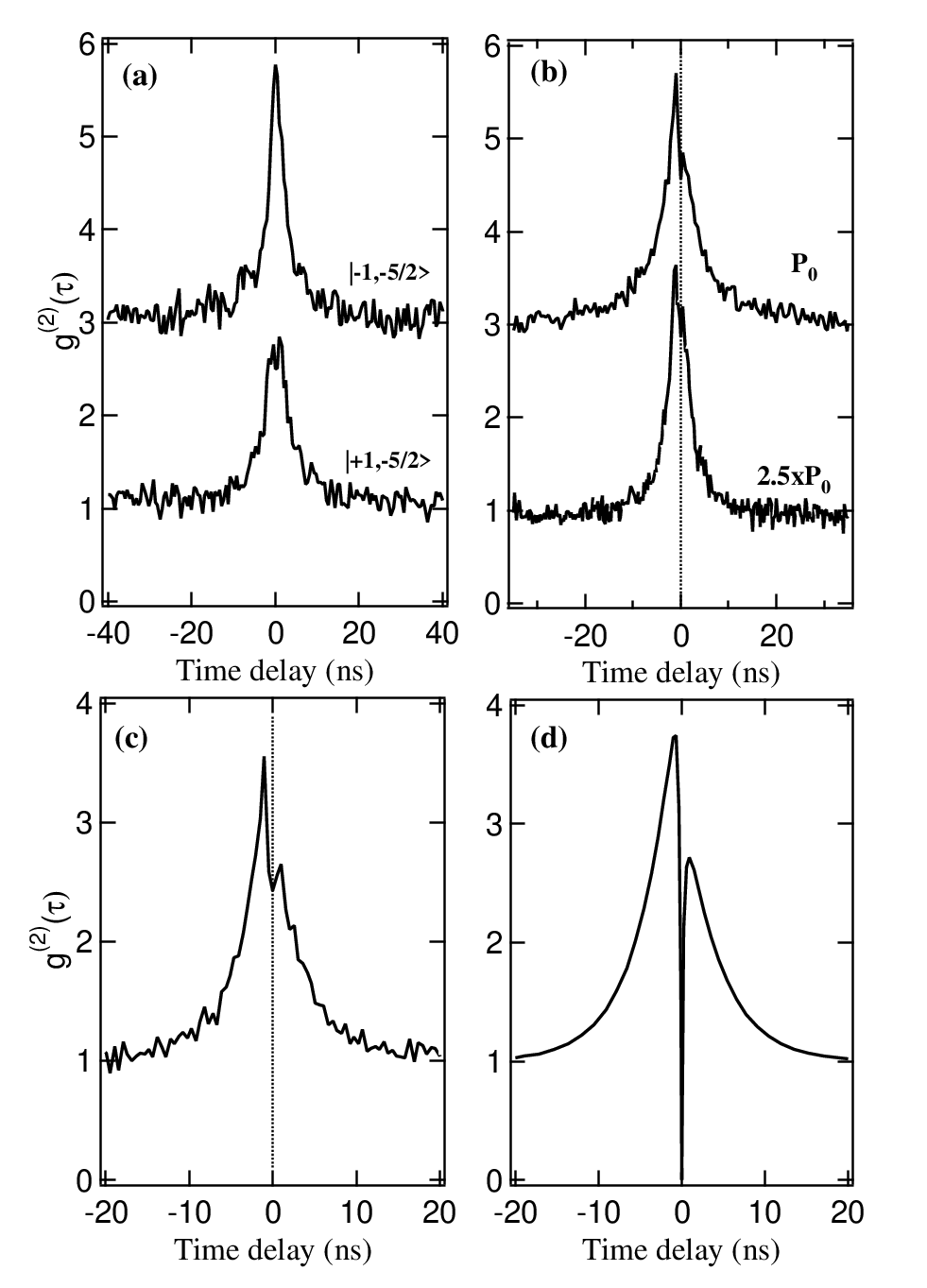}
\caption{ (a) Auto-correlation function of the emission intensity of
the hight (upper trace) and low (lower trace) energy lines of a X-Mn
complex recorded in the same circular polarization. (b)
Cross-correlation function of the emission intensity of the high and
low energy line recorded in $\sigma+$ and $\sigma-$ polarization
respectively for two different excitation intensities. Detail of the
experimental (c) and calculated (d) cross-correlation function.}
\label{fig5}
\end{figure}

For an isolated Mn atom, the spin relaxation $\tau_{Mn}$ comes only
from the spin-lattice interaction \cite{scalbert88} and a long spin
relaxation time is expected. As we will discuss in the next section,
the Mn spin dynamics can be modified significantly by the presence
of free carriers which are strongly coupled with both the magnetic
atom and the phonons. These free carriers serve as a bypass channel
for the slow direct spin-lattice relaxation.

\section{Thermalization of the exciton-Mn complex.}

The X-Mn complex is also significantly coupled to the phonon bath. A
partial thermalization of the X-Mn system appears directly in the
amplitude of the correlation curves obtained on different energy
levels of the X-Mn system (Fig.5(a)) as well as in cross-correlation
measurements (Fig.5(b) and (c)). A finite temperature enhances the
probability of the spin-flips involving an energy loss. This
introduces a dissymmetry in the spin relaxation channels of the X-Mn
complex. The consequence of this dissymmetry in the relaxation
process is an energy dependence of the amplitude of the correlation
signal. This is illustrated by the correlation curves obtained on
the high ($|-1,-5/2\rangle$) and low energy states
($|+1,-5/2\rangle$) of a X-Mn complex (Fig.5(a)). For the high
energy state, all the relaxation transitions within the X-Mn complex
take place with an energy loss: The leakage probability is then
maximum and the probability for this state to be re-populated by
spin-flips from low energy states is very weak. A large bunching
signal is then observed (Fig.5(a)). On the opposite, the low energy
level can be populated by a transfer from the high energy states,
and some relaxation channels involving an absorption of energy are
blocked at low temperature. The associated bunching signal is weaker
(Fig.5(a)).

This thermalization process directly appears if a cross-correlation
of the intensity emitted by a low and a high energy levels is
performed. Fig.5(c) shows the correlation of the photons emitted by
$|-1,-5/2\rangle$ (high energy line) and $|+1,-5/2\rangle$ (low
energy line). At low excitation intensity, this correlation signal
presents a clear dissymmetry. This cross-correlation measurement
probe the time dependence of the probability of finding an exciton
(either $|+1\rangle$ or $|-1\rangle$) coupled with the Mn spin in
the state M$_z$=-5/2. At positive time delay, g$^2(\tau)$ gives the
probability to find the system in the state $|+1,-5/2\rangle$
knowing that at $\tau=0$ the Mn spin projection was M$_z$=-5/2
(detection of a photon from $|-1,-5/2\rangle$). The situation is
reversed for the negative delay where a photon from
$|+1,-5/2\rangle$ acts as the trigger in the start-stop measurement
and g$^2(\tau)$ give the probability for the system to be in the
high energy state $|-1,-5/2\rangle$. The dissymmetry in the
cross-correlation curve reflects the difference in the spin
relaxation channels available for the high ($|-1,-5/2\rangle$) and
the low ($|+1,-5/2\rangle$) energy X-Mn states.

\begin{figure}[bt]
\includegraphics[width=2.8in]{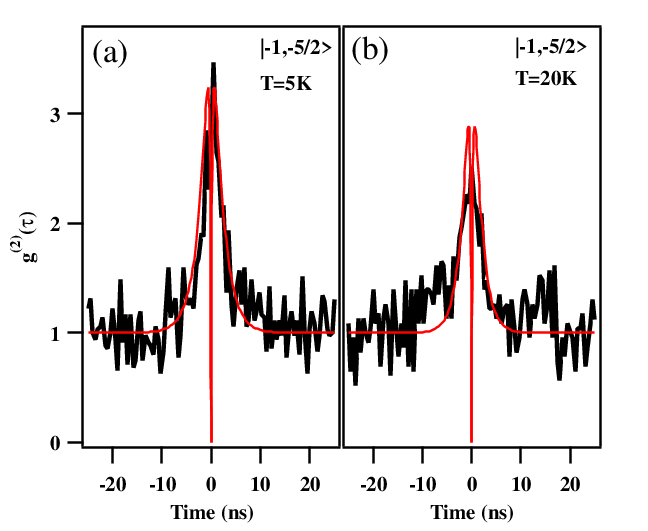}
\caption{ Intensity auto-correlation function of the high energy
line $|-1,-5/2\rangle$ of a X-Mn complex recorded in the same
excitation conditions at T=5K (a) and T=20K (b). The theoretical
curves (solid line) are obtained with the same set of parameters:
$\tau_{X-Mn}$=10ns, $\tau_{Mn}$=15ns,$\tau_{b}$=280ps,
$\tau_{d}$=8ns and g=0.1.10$^{-3}$ps$^{-1}$.} \label{fig6}
\end{figure}

The dissymmetry in the relaxation processes is influenced by the
excitation intensity: as observed in the PL decay measurements
presented in Fig.1, the low energy bright exciton states can be
efficiently populated by spin-flips from the dark exciton states
reducing the effective population loss of these states and
consequently reducing the amplitude of the photon bunching.
Increasing the excitation intensity decreases the effective lifetime
of the dark excitons because of the formation of the biexciton
\cite{besombes05}. This opens an efficient spin relaxation channel
for the low energy bright X-Mn states: once a dark exciton has been
created after a spin-flip, it is quickly destroyed by the injection
of a second exciton with the formation of a biexciton. It can no
longer flip back to the low energy bright state. This effect stop
the refilling process and consequently increases the amplitude of
the bunching signal. As observed in Fig.5(b), increasing the
excitation intensity decreases the difference in the amplitude of
the corresponding bunching signal of the low and high energy lines.

As shown in Fig.6, these spin relaxation processes are also
influenced by the lattice temperature. The effect is especially
pronounced for the high energy states. Increasing the temperature
allows a refilling of these levels by a transfer of population from
the low energy states, decreasing the amplitude of the bunching
signal.

\section{Model for the Mn Spin relaxation}

In this section we discuss two Mn spin  relaxation mechanisms   in
the optical ground state. The short ($\tau_{Mn}\simeq 20$ns) spin
relaxation time inferred from our photon correlation measurements
can not be accounted for by Mn-Mn spin interaction,   which is
considered the most efficient spin relaxation mechanism in II-VI
semiconductor. Since  this is  a short range interaction
\cite{Furdyna}, the Mn spin relaxation time, $T_{1M}$ increases
exponentially as the density of Mn goes down \cite{dietl95}. Thus,
the spin relaxation of a single Mn atom inside a quantum dot with 20
10$^3$ atoms, if generated by Mn-Mn coupling,  should be in the
range of microseconds. Thus,  some other spin relaxation mechanism
is at play. Here we study the Mn spin relaxation due to  exchange
coupling to photo-carriers that occupy extended states in the
wetting layer (WL). The Mn  is assumed to lie in the wetting layer,
at a location where both quantum dot and extended states are finite.
In particular, the QD states are expected to peak in the plane that
separates the wetting layer and the QD.  Mn atoms located in that
region are more strongly coupled to the QD exciton.

We consider two independent mechanisms: (i)exchange coupling with
carriers in extended WL states and (ii) exchange coupling with
carriers that scatter from the extended WL states to confined QD
states, exchange with the Mn and then return to the WL.  The first
mechanism has been considered before in the context of Mn spin
relaxation interacting with quantum well carriers
\cite{konig2000,Fonseca04} and is identical to Korringa mechanism
for nuclear spin relaxation due to spin-flip with itinerant
electrons in metals. The second mechanism involves single charge
tunneling in and out of the dot. The  QD confined states  remains
empty except for some  intervals during which a single carrier
tunnels in and out from the optically excited wetting layer. Once in
the QD state, the carrier can give (or take away)  one unit of spin
to  Mn. Both  the Korringa  and the charge fluctuation mechanisms
require several tunneling and exchange events, involving different
carriers, to relax the Mn spin from an initial situation where is
known to take the value $M_z=+5/2$ to a final situation of thermal
equilibrium, where the 6 Mn spin orientations are equally likely.

The  proposed charge fluctuation induced  spin relaxation mechanism
is in line with our previous
works\cite{Besombes04,JFR04,JFR06,Leger06,Fernandez2007}  whose
central claim is the very strong dependence of the spin properties
of Mn doped quantum dots on the addition of a single carrier. It is
also motivated by the observation of a weak contribution of the
positive trion PL signal which implies that, under our experimental
conditions, holes are captured by the QD.

\subsection{Formalism}

We use a Bloch-Redfield master equation approach\cite{cohen} which
tracks the dynamical evolution of the density matrix of the  quantum
dot ground states, corresponding to the 6 Mn spin orientations,
under the influence of the reservoir of carriers.  The diagonal
terms of the density matrix represent the probability of finding the
Mn spin in a given spin projection. Their dynammics is given by the
master equation
\begin{equation}
\frac{d p_M}{dt}= -\left(\sum_{M'} \Gamma_{M\rightarrow
M'}\right)p_M + \sum_{M} \Gamma_{M'\rightarrow M} p_{M'}
\end{equation}
with the initial condition that, at $t=0$, the probability of
finding the Mn spin with $M_z=+5/2$ is $p_{+5/2}=1$. The evolution
of $p_{+5/2}(t)$ and $p_{-5/2}(t)$ are directly associated to the
auto-correlation and cross-correlation measurements. The rates in
the master equation depend on  the Hamiltonian of the sytem, which
is the sum of the  QD Hamiltonian, the WL carrier Hamiltonian and
the QD-WL  coupling:
${\cal H}= {\cal H}_{0} + {\cal H}_{res} + {\cal V}$

$\Gamma_{M\rightarrow M'}$ is the scattering rate between the
eigenstate $M$ and  state $M'$ of ${\cal H}_0$ induced by the
coupling ${\cal V}$.  The rates are given by the statistical average
over initial reservoir states and sum over final reservoir states of
the Fermi Golden rule rate associated to ${\cal V}$ \cite{cohen}.
They depend on the mechanism under consideration, either direct
exchange or charge fluctuation. Since our experimental data indicate
that hole confinement is much weaker than electron confinement, we
consider that the carriers that couple to the Mn when there is no
exciton in the dot are  holes. Our theory results can easily be
adapted for the case of electrons. The number of states $M$ in the
master equation also depends on the mechanism. In the case of charge
fluctuation mechanism we need to keep track of the 6 neutral states
and the 12 states of the dot with one hole inside.  In the case of
the Korringa relaxation, only the 6  states corresponding to the
"empty dot" Mn spin are included in the master equation.

The QD  Hamiltonian  is:
\begin{equation}
{\cal H}_{0}=  J  \left(\tau^z {\bf \hat  M}^z +
\frac{\epsilon_h}{2}\left( \tau^{+}{\bf \hat  M}^{-} +\tau^{-}{\bf
\hat  M}^{+}\right)\right) +D \left({\bf \hat  M}^z\right)^2
\label{H_0}
\end{equation}
where  $J=\beta |\psi_{QD}(\vec{r}_I)|^2n_d$ is the Mn QD hole
coupling,
 $\beta\simeq 60 eV\AA^3$ is the Mn-hole exchange coupling constant of CdTe,
$|\psi_{QD}(\vec{r})|^2$ is the QD hole envelope function,
  $n_d=0,1$ counts the number of holes in the dot. The
Mn spin operators are  ${\bf \hat  M}^{z,\pm}$  and $\tau^{z,\pm}$
are the Pauli matrices acting on the  hole space. We consider
antiferromagnetic hole-Mn  coupling.
 $\epsilon_h$ is the  dimensionless parameter
that accounts for the reduction of the spin-flip interaction due to
spin-orbit coupling\cite{JFR06,Fernandez2007}.

Notice that the exchange coupling of the Mn to the quantum dot
fermion is only relevant in the charge-fluctuation mechanism, for
which  $n_d$ changes between 0 and 1. For the Korringa mechanism
$n_d=0$ and Mn is only exchanged coupled to carriers in extended WL
states.  The  $D ({\bf \hat M}^z)^2$ term in ${\cal H}_{QD0}$
describes the strain induced anisotropy which has been observed in
strained (Cd,Mn)Te layers \cite{Furdyna95}. This term is negligible
when the Mn interacts with a QD hole but is not when we consider the
Korringa relaxation, for which no carrier occupies the QD state. The
term  ${\cal H}_{\rm res}=\sum_{{\bf k},\nu} \epsilon_{{\bf k},\nu}
c^{\dagger}_{{\bf k},\nu}c_{{\bf k},\nu}$ describes the delocalized
carriers in states momentum   ${\bf k}$ and band index (including
the spin)  $\nu$.

Each of the two relaxation mechanisms considered here has its own
dot-reservoir coupling. In the Korringa mechanism we assume that
there is some overlap between the WL extended states and the Mn
spin. Igoniring the dependence on ${\bf k}$ of the spin matrix
element $\epsilon({\bf k},{\bf k}')$,the coupling reads:
\begin{equation}
{\cal V}\simeq \frac{\beta\epsilon}{2A}  |\phi(z_I)|^2 \sum_{\bf
k,k'}\left({\bf \hat M}^{(+)}c^{\dagger}_{{\bf k},\Downarrow}c_{{\bf
k}',\Uparrow} +
 {\bf \hat M}^{(-)}c^{\dagger}_{{\bf k}¡,\Uparrow}c_{{\bf k},\Downarrow} \right)
\end{equation}
where $A$  is the area of the WL, $|\phi(z_I)|^2$  is the envelope
part of the WL wave function evaluated at the Mn location,  and $L$
is the WL width. $\epsilon_h$ could take different values for  the
QD and the WL.

In the charge-fluctuation mechanism  the coupling between the
reservoir and the dot is the sum of an operator that transfers one
hole  from the reservoir to the dot and its hermitian conjugate,
which takes the hole out from the dot and transfers it to the
reservoir:
\begin{equation}
{\cal V}= \sum_{\sigma} |\sigma\rangle\langle0| \sum_{{\bf k}}
V_{{\bf k}} c_{{\bf k},\sigma} + h.c. \label{tunnel}
\end{equation}
The tunneling operator conserves the energy and the spin of the
carrier. This kind of coupling has been considered recently to
account for peculiar PL lineshapes of  self assembled quantum dots
in contact with electronic reservoirs \cite{Petroff05,Petroff08}.
The mechanism would operate analogously if the transfer from the
itinerant to the localized state is inelastic, but the timescale
would be longer.

\subsection{Results}
\subsubsection{ Mn spin relaxation due to exchange with WL carriers}
We discuss  first the relaxation of the Mn spin due to exchange with
WL carriers,  or Korringa mechanism. We assume a parabolic
dispersion with effective mass $m^*$ for the holes which yields a
stepwise density of states. We assume holes are in a thermal state
with effective temperature $k_BT_h$ larger than the lattice and
chemical potential $\mu$. The Korringa relaxation rate is
proportional to the square of the density of states at the Fermi
energy and the square of the exchange coupling constant. After some
work we obtain the  Mn spin flip rates  between $M$ and $M'$,
eigenstates of ${\cal H}_0=D({\bf \hat M}^z)^2$,   :
\begin{eqnarray}
\Gamma^{\pm}_{M\rightarrow M'}=
 \left(\frac{ \beta \epsilon  \eta m^*}{ \pi L\hbar^2  }\right)^2
  I(y,z) |\langle M|{\bf \hat M}^{(\pm)}|M'\rangle|^2
\end{eqnarray}
where we have approximated $|\phi(z_I)|^2= 2\eta/L$ , $0<\eta<1$ is
a dimensionless parameter that accounts for the overalp of the WL
states with the Mn, $y=\beta \mu$ is the chemical potential of the
WL carriers,  $z=(E_M-E_M')/k_BT_h$ is the change in energy of the
Mn spin divided by the effective temperature  of the WL carriers,
and $I(y,z)=k_BT_h
\frac{e^y}{e^{|z|}-1}Log\left[\frac{1+e^y}{1+e^{y-|z|}}\right]$ for
$z>0$ and $I(y,z)=e^{|z|}I(y,-z)$ for $z<0$.

\begin{figure}
[hbt]
\includegraphics[width=2.8in]{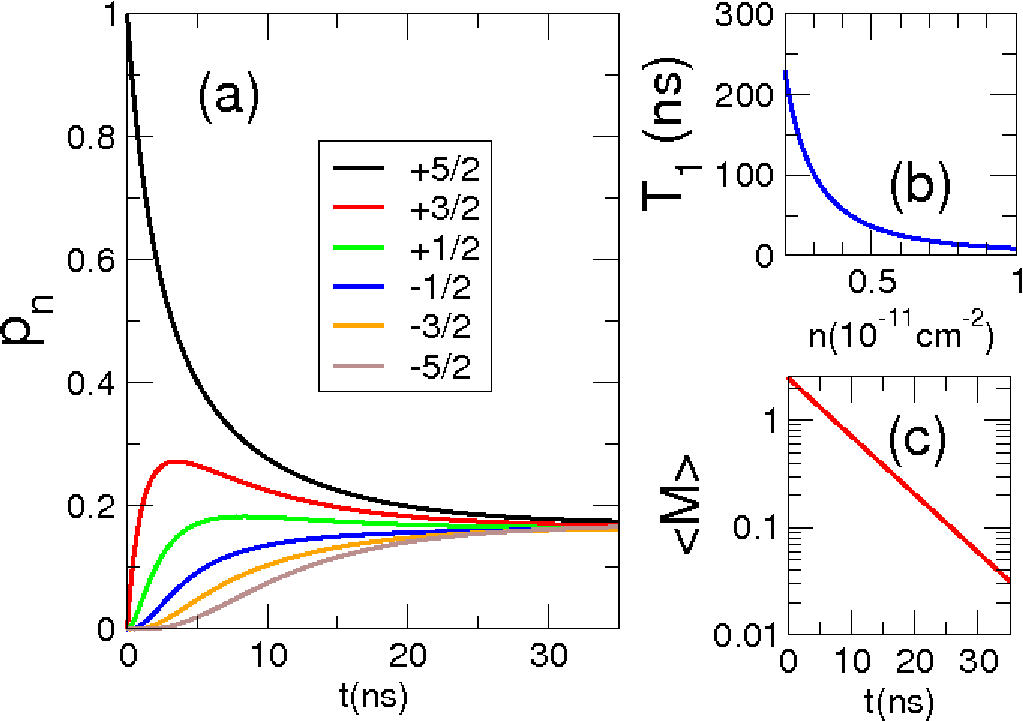}
\caption{ \label{relaxKO}(Color online). (a)  $p_n$ for Mn spin
states. These  can be considered as the {\em conditional
probabilities} that, having observed the Mn spin at $t=0$ in the
state $M_z=+5/2$,  a state with spin $M_z$ is observed at time $t$.
(b) $T_{1M}$ as a function of hole density (see text). (c) Average
magnetization corresponding to panel (a), in logarithmic scale.}
\end{figure}

The solution of the master equation, taking as initial condition
$p_{+5/2}=1$, is shown in figure (\ref{relaxKO}). In the left panel
we plot the evolution of the occupations of the 6 Mn spin
projections along the growth axis.  The decay of $p_{+5/2}$ is
accompained by a  rise of the $+3/2$ state. This state is also
connected with the $1/2$ state whose occupation starts to build.
Interestingly, the population of the $+3/2$ states overshoots
initially. The figure shows a "falling domino" effect. They evolve
towards thermal equilibrium. The average magnetization $\langle M
\rangle(t) = \sum_{n=1,6} M_n p_n(t)$ decays according to
\begin{equation}
\langle M \rangle(t) =M_0 e^{-t/T_{1M}}
\end{equation}
 in all the simulations performed.
In the figure we take $T_h\simeq 10 K$, $\epsilon=0.1$, $\eta=1$,
${\cal D}=7\mu$eV, $m^*$=0.4, $L=20 \AA$,  and $n_h=10^{11}cm^{-2}$
and we obtain $T_{1M}\simeq 8$ ns. Thus, this mechanism is
consistent with the Mn spin relaxation time that we observe. In the
figure (\ref{relaxKO}) we see how $T_{1M}$ increases as the density
of carriers decreases. The order of magnitude of $T_{1M}$ is
consistent with the calculations for Mn spin relaxation induced by
scattering with electrons in quantum wells \cite{konig2000}. There
$T_{1M}\simeq 2\times 10^3$ ns is obtained in a quantum well with
$L=80\AA$. Taking into account that the electron Mn coupling
$\alpha$ is 4 times smaller than $\beta$ and the $L$ is 4 times
bigger, there $T_{1M}$ for holes should be at least  $16^2$ shorter.

Thus Korringa exchange with a sufficiently large density of
photoholes might account for the Mn spin relaxation when the dot is
in the optical ground state. We consider now a different mechanism,
involving single hole charging of the dot,  motivated by the fact
that  the PL often shows a weak contribution of  positive trions.

\subsubsection{ Mn spin relaxation due to charge fluctuation in the dot}

The elementary process for spin relaxation due to charge fluctuation
is a combination of 3 steps. First, a hole tunnels into the empty
dot with the Mn in the spin state $M$. Second, the hole and the Mn
exchange spin, so that the Mn is now in the state $M\pm1$. Third,
the hole tunnels out of the dot. Thus, in the master equation we
need to include both the 6 charge neutral states of the Mn doped dot
as well as the 12  states with one hole inside. The dissipative
dynamics, induced by ${\cal V}$ connects neutral states with charged
states ($+$)  and viceversa ($-$). The charging transition rates are
given by
\begin{equation}
\Gamma^{(+)}_{n,m}= \Gamma_0 n_F\left((E_n-E_m)\right) |\langle
n|\sum_{\sigma} |0\rangle\langle \sigma ||m\rangle |^2
\label{rateplus}
\end{equation}
and the decharging rate:
\begin{equation}
\Gamma^{(-)}_{n,m}= \Gamma_0 \left(1-n_F(E_n-E_m)\right) |\langle
n|\sum_{\sigma} |0\rangle\langle \sigma ||m\rangle |^2
\label{rateminus}
\end{equation}
Here $n_F(x)$ is the Fermi function which depends on the temperature
and chemical potential of the electron reservoir, $E_n$ and $E_m$
are the eigenstates of the QD Hamiltonian  and $\Gamma_0=\sum_k
|V_k|^2 \rho(E_f)$ is the tunneling rate for the fermion in and out
of the dot. The matrix elements featured in eq.
(\ref{rateplus},\ref{rateminus}) are strongly spin dependent and
only connect states in which the Mn spin changes, at most, by one
unit.

An important quantity in our simulations is the average charge,
$\langle q \rangle =\sum_{n=7,18} p_n(t)$. Since we consider an
initally neutral dot, this quantity goes from zero to the steady
state occupation in a time scale given by
$T_{1Q}\equiv(\Gamma_0)^{-1}$. In this second set of simulations  we
measure time in units of $T_{1Q}$. The evolution of the
magnetization and the empty dot occupations on this case look very
similar to those of figure (\ref{relaxKO}). In particular, the
magnetization decays exponentially in a time scale much longer than
the charge relaxation time.

\begin{figure}
[hbt]
\includegraphics[width=2.8in]{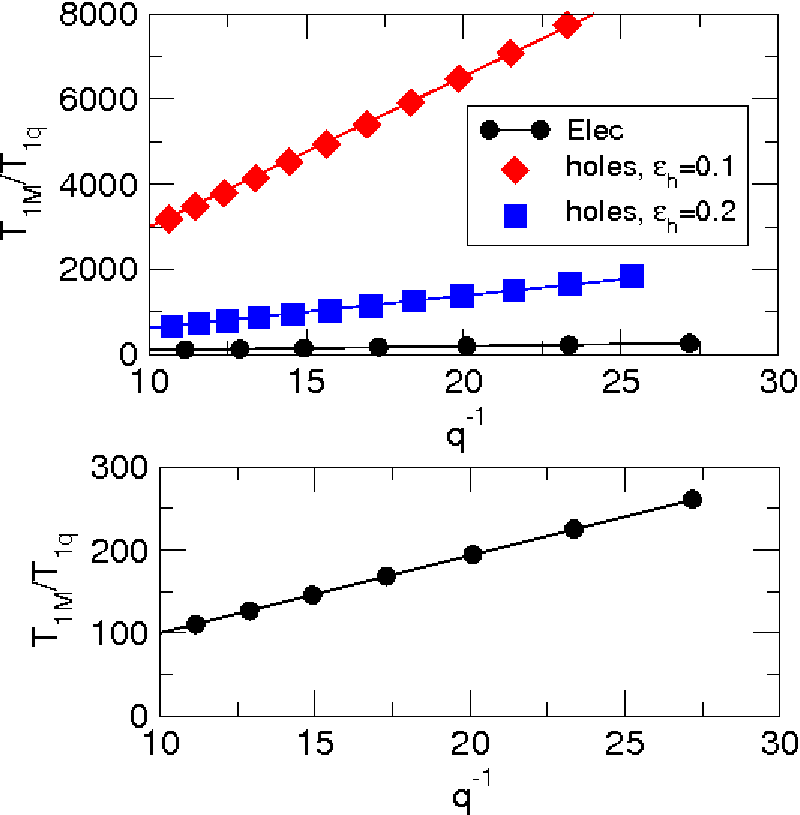}
\caption{ \label{summary}(Color online). Mn spin relaxation time,
originated by charge fluctuations in the dot, as a function of
average charge in the dot for electrons and holes with different
$\epsilon_h$ values. The straight lines are linear fits of the
numerical data.}
\end{figure}

We have computed the Mn spin relaxation time, $T_{1M}$, as a
function of the parameters of the simulation, $\Gamma_0$, $k_B T$,
$J$, $\epsilon$  and $\langle q\rangle$. We have found that the
crucial quantities are the average charge, the spin flip anisotropy
$\epsilon_h$ and $\Gamma_0$, as shown in figure (\ref{summary}). To
very good approximation, we have that $\Gamma_0 T_{1M}$ depends
linearly on $\langle q\rangle^{-1}$, for moderate values of $\langle
q\rangle$. Thus, this mechanism shows a strong dependence on the
density of carriers in the WL, in agreement with our observations.

In figure \ref{summary}  we see that, in the case of tunneling
electrons ($\epsilon=1$), for an average charge of $0.1$ we have
$T_{1M}\simeq 100 \Gamma_0 ^{-1}$.  For tunneling holes with
$\epsilon_h=0.2$ and $\langle q\rangle=0.1$ we have $T_{1M}\simeq
600 \Gamma_0^{-1}$. Interestingly, these numbers are weakly
dependent of the value of $J$ and  on the temperature. In order to
obtain an absolute number for $T^M_1$ we need an estimate of
$\Gamma_0$ the charge scattering rate. An upper limit for $\Gamma_0$
is provided by the linewidth observed in the PL spectrum. Since the
charge scattering would result in a broadening of the linewidth, we
can infer that $\hbar \Gamma_0< $ 50$\mu$eV. Thus, using
$\hbar=0.65$meV$\times$ps we have $T^M_1 > 600
\frac{\hbar}{\Gamma_0} \simeq 8 ns $ for $\langle q \rangle=0.1$.
This is a lower bound for the spin scattering time, or an upper
bound for the spin scattering rate.  Thus, single hole tunneling
events could relax the spin of the Mn in a time scale of 8ns for a
dot which is charged 10 percent of the time. Agreement with the
experimental result can be obtained by taking  smaller average
charge in the dot  or  smaller spin-flip interaction or smaller
tunneling rate $\Gamma_0$.  We note that the proposed mechanism is
similar to the hole spin-relaxation mechanism proposed by Smith {\em
et al.} \cite{Petroff05}. There, a spin up hole tunnels out of the
dot and spin down tunnels in, resulting in an effective spin
relaxation for the hole.

In summary, we have  two mechanisms that account for carrier induced
Mn relaxation in the dot in a time scale of 10 nanoseconds. In both
cases the dot must be considered as an open system. In the Korringa
relaxation mechanism the Mn spin is exposed to extended WL states.
In the charge-fluctuation mechanism  the occupation of the dot state
fluctuates, due to its coupling to extended WL states. These two
mechanisms are not exclusive and might operate at the same time.
Whereas the Korringa  mechanism is almost identical to that of
(Cd,Mn)Te quantum wells \cite{konig2000}, the charge-fluctuation
mechanism is specific of quantum dots coupled to a continuum.

\section{Conclusion.}

In conclusion, we used time resolved and photon correlation
measurements to probe the spin dynamics of a single magnetic atom
(Mn) interacting with photo-created carriers at zero magnetic field.
Fluctuations of the localized Mn spin control the statistics of the
photons emitted by a single Mn doped QD. The spin relaxation time of
the Mn atom, around 20ns, appears to be particularly sensitive to
the spin-spin coupling with free carriers injected in the QD or on
the vicinity of the QD. The fast photo-induced spin relaxation of
the Mn is dominated by spin and energy transfer from hot
photo-carriers directly to the Mn atom. Minimizing the injection of
free carriers by optimizing the excitation conditions would make it
possible to reach longer fluctuation times. As a consequence,a
determination of the intrinsic Mn spin dynamics at zero magnetic
field, ultimately limited by the spin-lattice interaction, could
only be performed in the absence of free or confined carriers.

This work was supported by French ANR contracts MOMES and CoSin.


\begin{thebibliography}{}

\bibitem{xiao2004}M. Xiao, I. Martin,E. Yablonovitch,H. W. Jiang, Nature (London) 430, 435
(2004).

\bibitem{Fernandez2007}J. Fernandez-Rossier and R. Aguado, Phys. Rev. Lett. 98, 106805 (2007)

\bibitem{lambe60} J. Lambe and C. Kikuchi, Physical Review, 119,
1256 (1960).

\bibitem{kneip06} M.K. Kneip, D. R. Yakovlev, M. Bayer, A. A. Maksimov, I. I. Tartakovskii, D. Keller,
W. Ossau, L. W. Molenkamp, A. Waag, Phys. Rev. B {\bf 73}, 045305
(2006)

\bibitem{dietl95} T. Dietl, P. Peyla, W. Grieshaber, Y. Merle d'Aubigne,  Phys. Rev. Lett. 74, 474 (1995).

\bibitem{heinrich2004} A.J. Heinrich, J.A. Gupta, C.P. Lutz, D.M. Eigler, Science 306, 466
(2004)

\bibitem{Kitchen2006} D. Kitchen, A. Richardella, J.-M. Tang, M. E. Flatte, and A. Yazdani, Nature (London) 442,
436 (2006).

\bibitem{Besombes04} L. Besombes, Y. L\'eger, L. Maingault, D. Ferrand, H. Mariette, J. Cibert, Phys. Rev. Lett. {\bf 93}, 207403 (2004)

\bibitem{Maingault2006} L. Maingault, L. Besombes, Y. L\'eger, C. Bougerol, H. Mariette , Appl. Phys. Lett. {\bf
89}, 193109  (2006).

\bibitem{Couteau2004} C. Couteau, S. Moehl, F. Tinjod, J.M. G\'erard, K. Kheng, H. Mariette, J. A. Gaj, R. Romestain, J.P. Poizat, Appl. Phys. Lett. {\bf
85}, 6251 (2004).

\bibitem{leger05} Y. L\'eger, L. Besombes, L. Maingault, D. Ferrand, H. Mariette, Phys. Rev. B {\bf 72}, 241309 (2005).

\bibitem{bes05}  L. Besombes ,Y. Leger, L. Maingault, D. Ferrand, H. Mariette, J. Cibert, Phys. Rev. B {\bf 73}, 161307 (2005).

\bibitem{smith05} J.M. Smith P. A. Dalgarno, R.J. Warburton, A.O. Govorov, K. Karrai, B. D. Gerardot, P. M. Petroff, Phys. Rev. Lett. {\bf 94}, 197402 (2005)

\bibitem{kiraz02} A. Kiraz, S. Fälth, C. Becher, B. Gayral, W.V. Schoenfeld, P. M. Petroff, Lidong Zhang, E. Hu, A. Imamoglu, Phys. Rev. B {\bf 65}, 161303(R) (2002)

\bibitem{scalbert88} D. Scalbert, J. Cernogora, C. Benoit à la Guillaume, Solid State Commun.{\bf 66}, 571 (1988).

\bibitem{govorov2005} A. O. Govorov, A. V. Kalameitsev, Phys. Rev. B {\bf 71}, 035338 (2005).

\bibitem{feng2007} D.H. Feng, I.A. Akimov, F.Henneberger, Phys. Rev. Lett. {\bf99}, 036604 (2007).

\bibitem{vasanelli02} A. Vasanelli, R. Ferreira, and G. Bastard, Phys. Rev. Lett. {\bf89}, 216804 (2002)

\bibitem{Tyazhlov97} M.G. Tyazhlov, A.I. Fillin, A.V. Larionov, V.D.
Kulakovskii, D.R. Yakovlev, A. Waag, G. Landwehr, JETP {\bf 85}, 784
(1997).

\bibitem{besombes05} L. Besombes, Y.Leger, L. Maingault, D.Ferrand, H.Mariette, Phys. Rev. B {\bf 71}, 161307(R) (2005).

\bibitem{woods2004} L.M. Woods, T. L. Reinecke, and R. Kotlyar, Phys. Rev. B {\bf 69}, 125330 (2004).

\bibitem{Leger2007} Y. L\'eger, L. Besombes, L. Maingault, H. Mariette, Phys. Rev. B {\bf76}, 045331 (2007).


\bibitem{Furdyna} J. K. Furdyna, J. Appl. Phys {\bf 64} R29 (1988).


\bibitem{T1} J. Lambe and C. Kikuchi, Phys. Rev. {\bf 119}, 1256 (1960).
D. Scalbert {\em et al.}, Solid State Communications {\bf 66}, 571
(1988).

\bibitem{konig2000}B. Koenig, I. A. Merkulov, D. R. Yakovlev, W. Ossau, S. M.
Ryabchenko, M. Kutrowski, T. Wojtowicz, G. Karczewski, and J. Kossut
Phys. Rev. B {\bf 61}, 16870 (2000)

\bibitem{Fonseca04} E. Souto E, O. A. C. Nunes , A. L. A.  Fonseca, Solid State Comm.
{\bf 129}, 605 (2004)

\bibitem{Leger06} Y. L\'eger
 L. Besombes, J. Fern\'andez-Rossier,
 L. Maingault, H. Mariette , Phys. Rev. Lett.{\bf 97},  (2006)

\bibitem{JFR06} J. Fern\'andez-Rossier,
Phys. Rev. B.{\bf 73}, 045301 (2006)

\bibitem{JFR04} J. Fern\'andez-Rossier and
 L. Brey, Phys. Rev. Lett.  {\bf 93} 117201 (2004)

\bibitem{cohen} Claude Cohen-Tannoudji, Jacques Dupont-Roc, Gilbert Grynberg
{\em Atom-Photon Interactions: Basic Processes and Applications},
John Wiley and Sons, New York, 1992.

\bibitem{Furdyna95} M. Qazzaz, G. Yang, S. H. Xin, L. Montes, H. Luo, J. K. Furdyna
Solid State Comm. {\bf  96},  405 (1995)

\bibitem{Petroff05}
J. M. Smith, P. A. Dalgarno, R. J. Warburton, A. O. Govorov, K.
Karrai, B. D. Gerardot, and P. M. Petroff Phys. Rev. Lett. {\bf 94},
197402 (2005)

\bibitem{Petroff08}
 P. A. Dalgarno, M. Ediger, B. D. Gerardot, J. M. Smith, S. Seidl, M. Kroner, K. Karrai, P. M. Petroff, A. O. Govorov, and R. J. Warburton
Phys. Rev. Lett. {\bf 100}, 176801 (2008)


\end{thebibliography}
\end{document}